\documentclass[aps,pra,twocolumn]{revtex4-1}
\usepackage{braket}
\usepackage{amsmath,bbm}
\usepackage{amssymb}
\usepackage{graphicx}
\usepackage{amsthm}

\newcommand{\beq}{\begin{equation}}
\newcommand{\eeq}{\end{equation}}
\newcommand{\beqa}{\begin{eqnarray}}
\newcommand{\eeqa}{\end{eqnarray}}

\def\opone{\leavevmode\hbox{\small1\normalsize\kern-.33em1}}

\graphicspath{ {../Figs/} }

\begin{document}

\renewcommand{\today}{\number\day\space\ifcase\month\or
   January\or February\or March\or April\or May\or June\or
   July\or August\or September\or October\or November\or December\fi
   \space\number\year}

\begin{center}
\title{Macroscopic Quantum Measurements of noncommuting observables}

\date{\today}

\author{Tomer~Jack~Barnea}
\thanks{The first two authors contributed equally to this work.}
\author{Marc-Olivier~Renou}
\thanks{The first two authors contributed equally to this work.}
\author{Florian Fr{\"o}wis}
\email{Electronic address: florian.froewis@unige.ch}
\author{Nicolas~Gisin}
\affiliation{Group of Applied Physics, University of Geneva, 1211 Geneva 4, Switzerland}

\begin{abstract}
  Assuming a well-behaving quantum-to-classical transition, measuring large quantum systems should be highly informative with low measurement-induced disturbance, while the coupling between system and measurement apparatus is ``fairly simple'' and weak. Here, we show that this is indeed possible within the formalism of quantum mechanics. We discuss an example of estimating the collective magnetization of a spin ensemble by simultaneous measuring three orthogonal spin directions. For the task of estimating the direction of a spin-coherent state, we find that the average guessing fidelity and the system disturbance are nonmonotonic functions of the coupling strength. Strikingly, we discover an intermediate regime for the coupling strength where the guessing fidelity is quasi-optimal, while the measured state is almost not disturbed. 
\end{abstract}
\maketitle
\end{center}

\section{Introduction}
\label{sec:introduction}

In everyday life we continuously perform measurements. For instance, to locate our friends we perform some kind of position measurements; similarly to read this text. Presumably all this can be described with the quantum formalism. But, obviously, these measurements are not of the standard von Neumann projective kind. 
They are highly noninvasive while still collecting a large amount of information in a global, single shot. Additionally, from a physical point of view, we expect a fairly simple coupling between system and observer. We call measurements that fulfill these requirements ``macroscopic quantum measurements''. Our goal is to see if and how macroscopic quantum measurement can be realized. 

For concreteness, think of a magnet whose $N$ atoms are idealized by spin 1/2 particles and assume that all spins are aligned parallel to an unknown direction $\vec{u} \in \mathbb{R}$. By a freely selectable measurement, we are asked to estimate $\vec{u}$. We define the score function $\cos^{2} \theta/2$, where $\theta$ is the angle between $\vec{u}$ and the best guess $\vec{w}$. This task can be seen as a simultaneous measurement of the magnetization in the three principal directions $x, y$ and $z$. The corresponding measurement operators are collective spin operators $S_x = \frac{1}{2} \sum_i \sigma_x^{(i)}$ (and similarly for $y$ and $z$). The maximal average guessing fidelity is $\mathcal{F}_{\mathrm{opt}} = (N+1)/(N+2) = 1 - 1/N + O(N^{-2})$, which is achievable either with nontrivial, entangling projective measurements or with a continuous general measurement with elements $\Omega(\vec{r}) \propto \left| \vec{r} \right\rangle\!\left\langle \vec{r}\right| ^{\otimes N}$ \cite{massar95}. The same fidelity can be asymptotically reached by random measurements on individual spins \cite{bagan05}.

Here, in order to realize the simultaneous measurement of $x$, $y$ and $z$, we build upon von Neumann's measurement model by coupling all three observables $S_x$, $S_y$ and $S_z$ to three independent pointers. Each pointer is then individually measured and the results are processed to guess $\vec{w}$ \cite{dariano02}. (This is in the spirit of the Arthurs-Kelly model \cite{arthurs65}, where position and momentum are coupled to two individual pointers.) In the strong coupling regime, this measurement corresponds to randomly choosing a direction $\vec{n}$ and to strongly measure along this axis \cite{dariano02}. This measurement model leads to an average fidelity of 3/4 (i.e., far away from the optimal value) for large $N$ \footnote{The average fidelity of 3/4 is exactly the same value as estimating a classical spin with a Stern-Gerlach-type experiment with randomly chosen measurement axis.}.

\begin{figure}[b]
  \centerline{\includegraphics[width=\columnwidth]{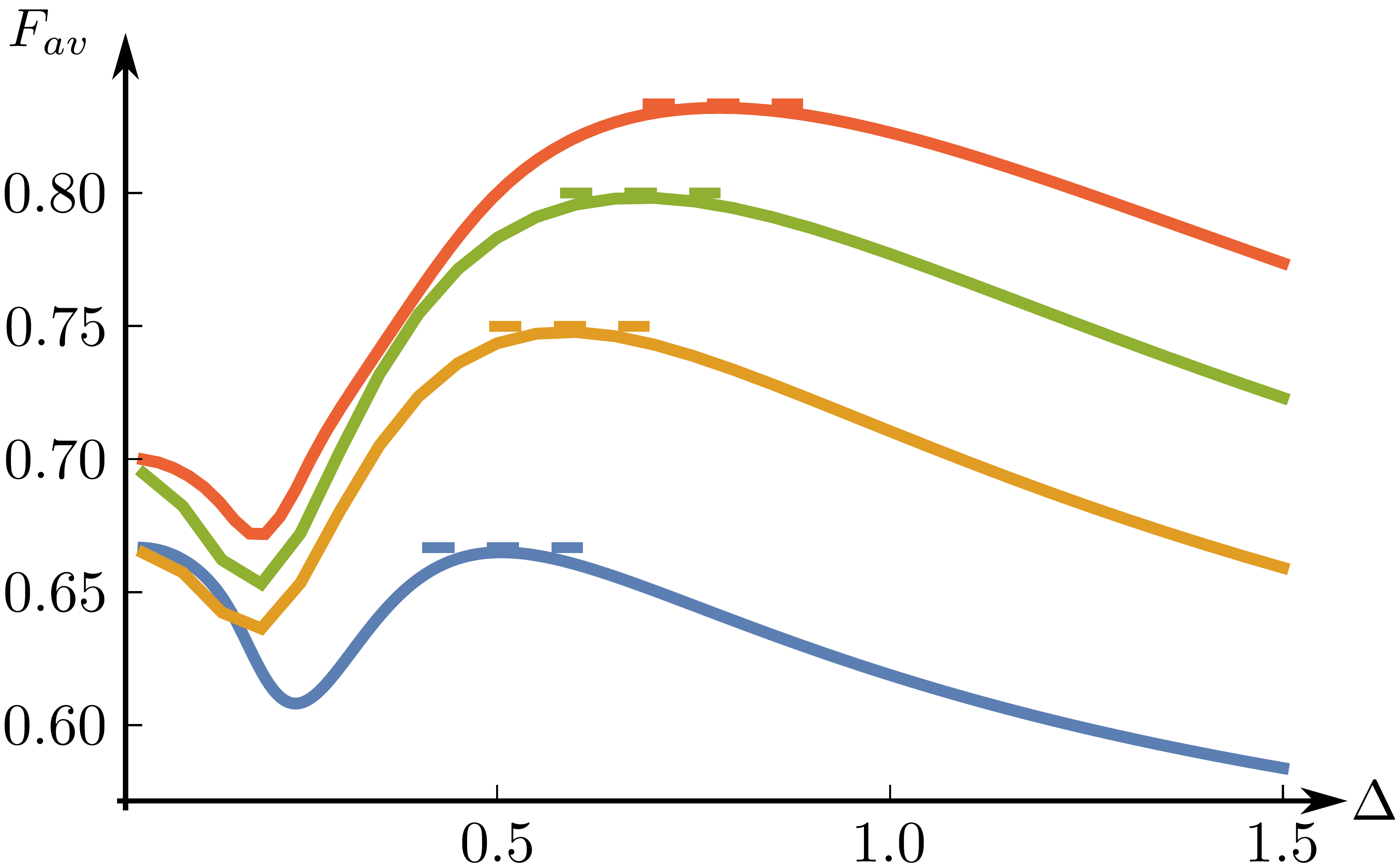}}
  \caption[]{\label{fig:avfids} Average fidelity $\mathcal{F}_{\mathrm{av}}$ for guessing the unknown state as a function of the inverse coupling strength $\Delta$ for $N = 1,\dots,4$ (from bottom to top). The dashed lines correspond to the optimal value $\mathcal{F}_{\mathrm{opt}}$. For $N\geq 2$, the global maximum is reached by some nonzero $\Delta$ and is close to the optimal value.}
\end{figure}

The coupling strength between system and pointers influences the guessing fidelity and the measurement-induced disturbance. As shown by Poulin \cite{Poulin_Macroscopic_2005}, the measurement of a single observable (e.g., only $S_x$) is more informative and more disturbing as the coupling strength increases for a natural choice of the initial pointer state and the final measurement. For $N$ parallel spins measured in any direction, it turns out that the system is asymptotically undisturbed if the coupling is much less than $N^{-1/2}$ \cite{Poulin_Macroscopic_2005}. Poulin then suggested to perform many measurements in this weak coupling regime; potentially for several noncommuting observables. He conjectured that this should eventually give maximal information about the initial state without disturbing it. However, apart from practical questions, the limit of information gain and the measurement-induced disturbance of ``infinitely many and infinitely weak'' measurements is unclear so far. Other contributions, like Ref.~\cite{Busshardt_Timing_2010}, discuss the influence of the coupling strength, but do not compare to optimal strategies.

In this paper, we present an example of the surprising nonmonotonic relation between information gain, measurement-induced disturbance and the coupling strength when noncommuting observables are simultaneously measured. By studying the example of estimating the direction of parallel spins, we find an intermediate regime for the coupling strength where the average guessing fidelity is numerically very close to the optimal value $\mathcal{F}_{\mathrm{opt}}$; both for small and large $N$. In the very same coupling regime, the measurement-induced disturbance is small, in particular as $N$ increases. More specifically, the averaged Bloch vector of the post-measured states is almost identical with the initial state, meaning that repeating the protocol by independent observers gives almost the same estimates. We claim that the quasi-optimal information gain and small measurement-induced disturbance is reminiscent of classical measurements. In this sense, the presented model is an example of a quantum measurement that behaves classically for large system sizes.

\section{General measurement model and average fidelity}
\label{sec:measurement-model}
In the following, we denote a normalized spin state orientated in the direction of a general vector $\vec{x} \in \mathbbm{R}^3$ by $\ket{\vec{x}} \in \mathbbm{C}^2$. Generally, we consider a scheme where the system state $\ket{\vec{u}}^{\otimes N}  $ is coupled to a measurement apparatus $| \phi \rangle $ --called pointer-- through virtue of an interaction Hamiltonian $H_{\mathrm{int}}$. Afterwards, the pointer is measured. This procedure is often referred to as the von Neumann measurement model \cite{neumann55,Busch_Quantum_1991,peres02}. For simplicity, the coupling with strength $\mu$ is uniformly switched on for a unit time and dominates meanwhile all other processes; thus $U = \exp(-i \mu H_{\mathrm{int}})$. After this the joint state of system and pointer reads $U \ket{\phi} \otimes \ket{\vec{u}}^{\otimes N}$. The subsequent measurement of the pointer is modeled by a projective measurement with outcome $\vec{r}$ for the eigenvectors $| \psi_{\vec{r}} \rangle $. (The dimension of $\vec{r}$ depends on the pointer space.) This outcome is then classically processed. Knowledge about the initial pointer state, the coupling and the final measurement allows one to calculate the optimal guess state $\left|  \vec{w}_{\vec{r}}\right\rangle $.
So far, this is a fully general measurement procedure and is linked to the Kraus operator formalism via
\begin{equation}
\label{int}
E(\vec{r}) = \bra{\psi_{\vec{r}}} U \ket{\phi}.
\end{equation}

To measure the accuracy of the measurement procedure we choose the average fidelity $\left| \left\langle \vec{w}_{\vec{r}} \right| \left. \vec{u}\right\rangle  \right|^2$. The average score is then given by
\begin{equation}
\label{fid}
\mathcal{F}_{\mathrm{av}} = \int \frac{d\vec{u}}{4\pi}~\int d\vec{r}~p(\vec{r}|\vec{u})~\lvert\braket{\vec{w}_{\vec{r}}|\vec{u}}\rvert^2,
\end{equation}
where $p(\vec{r}|\vec{u}) = \lVert E(\vec{r}) \left| \vec{u} \right\rangle ^{\otimes N} \rVert^2$ is the probability to get outcome $\vec{r}$ given that the initial state was $\left|   \vec{u}\right\rangle^{\otimes N}$. 

A typical instance is a pointer with one spatial degree of freedom \cite{Poulin_Macroscopic_2005}. For example, the initial spatial wave function of the pointer is a Gaussian function with spread $\Delta$. In order to measure an observable $A$, one then defines $H_{\mathrm{int}} = p\otimes A$, where $p$ represents the displacement operator in the pointer space, which is formally equivalent to the momentum operator. Thus, the coupling induces a momentum kick on the pointer whose strength depends on the initial system state. Finally, a position measurement of the pointer allows some inference about the system. Information gain and invasiveness of this procedure manifest themselves in the relationship between the coupling strength $\mu$, the spread of the Gaussian, $\Delta$, and the spectrum of the eigenvalues of the operator $A$. By redefining $\Delta$, one can set $\mu = 1$ without loss of generality. Thus, the coupling strength of the measurement is proportional to $\Delta^{-1}$. A $\Delta$ small compared to the spectral gap of $A$ corresponds to a strong coupling that resolves the individual eigenvalues. The momentum kicks induced by two nearby eigenstates of $A$ are distinguishable. On the other side, a $\Delta$ large compared to the spectral gap of $A$ means that system states prepared in neighboring eigenstates cannot be well distinguished. This is the regime of fuzzy measurements \cite{peres02}. Note that the maximal information gain \textit{and} the maximal disturbance are typically realized in the limit of a vanishing $\Delta$, that is, when the pointer model approaches the ideal projective measurement of $A$. Hence, even for large $N$, the basic idea of small invasiveness and large information gain does not seem to be reachable with this simple construction.

\section{Specific modeling}
\label{sec:macr-quant-meas}

To measure the direction of $| \vec{u} \rangle^{\otimes N}$, a pointer with three spatial degrees of freedom seems to be more appropriate. We choose
\begin{equation}
\label{pointer}
\ket{\phi} = \frac{1}{(2 \pi \Delta^2)^{3/4}} \int{dx dy dz ~e^{-\frac{(x^2+y^2+z^2)}{4 \Delta^2}} \ket{x} \ket{y} \ket{z}},
\end{equation}
where $x$, $y$ and $z$ denote the coordinates of the pointer. The spatial distribution is a rotationally invariant Gaussian function with spread $\Delta$. The direction of $| \vec{u} \rangle^{\otimes N} $ is determined by the three expectation values of the collective spin operators $S_x, S_y$ and $S_z$. 
Thus, a classically inspired interaction Hamiltonian reads
\begin{equation}
\label{ham}
H_{\mathrm{int}} = p_x \otimes S_x + p_y \otimes S_y + p_z \otimes S_z = \vec{p}\cdot \vec{S},
\end{equation}
which follows the pattern of the Arthurs-Kelly model. 
The $p_k$ for $k = x,y,z$ represent the displacement operators in the $x$, $y$ and $z$ directions, respectively. Finally, a position measurement with outcome $\vec{r} \in \mathbbm{R}^3$ is performed on the pointer. We simply guess $| \vec{w}_{\vec{r}} \rangle = \left| \vec{r} \right\rangle $\footnote{Note that this choice is not always optimal. For some (nonoptimal) values of $\Delta$, the optimal choice is $| \vec w_{\vec r} \protect\rangle = | -\vec r \protect\rangle $. For symmetry reasons the optimal guess has to lie on the axis of $\vec r$.}. The model is rotationally symmetric. Without loss of generality, we hence consider a fixed input state $| \vec{u} \rangle = \left| \hat{e}_z \right\rangle \equiv \left| \uparrow \right\rangle $. With $p(\vec{r}) \equiv p(\vec{r}|\hat{e}_z )$, Eq.~\eqref{fid} then simplifies to 
\begin{equation}
\label{eq:11}
\mathcal{F}_{\mathrm{av}} = 2\pi\int r^2 \sin \theta dr d\theta  p(\vec{r}) \cos^2 \theta/2,
\end{equation}
since $p(\vec{r})$ does not depend on the azimuthal angle.

In the limit $\Delta \rightarrow 0$, the measurement model is suboptimal and the average fidelity is $\mathcal{F}_{\mathrm{av}} = \frac{1}{4} (3N+2)/(N+1)$ for even $N$ and  $\mathcal{F}_{\mathrm{av}} = \frac{1}{4} (3N+5)/(N+2)$ for odd $N$  \cite{dariano02}. These values are bounded by $\frac{3}{4}$ and are hence far away from the optimal fidelity. As we will show now, the three spatial dimensions of the pointer combined with noncommuting operators in Eq.~\eqref{ham} involve a rich behavior for nonvanishing $\Delta$. In contrast to the one-dimensional pointer, information gain and disturbance are not monotonically related to the coupling strength.

\section{Small number of spins}
\label{sec:small-number-spins} We start by calculating the average fidelity $\mathcal{F}_{\mathrm{av}}$ for one to four spins as a function of the spread $\Delta$ of the Gaussian pointer. In these cases the expression in Eq.~\eqref{eq:11} is easy to work with due to the small number of spins. The Kraus operators $E(\vec{r})$ of the interaction, Eq.~\eqref{int}, can be derived analytically. The average fidelity is analytically calculated in the case of one single qubit, while for more spins the integration over the radial component of $\vec{r}$ is performed numerically. The results are shown in Fig.~\ref{fig:avfids}. For two or more spins the optimal $\Delta$ is clearly distinct from zero and $\mathcal{F}_{\mathrm{opt}}$ (dashed lines in Fig.~\ref{fig:avfids}) can almost be reached.

Analyzing the graph shown in Fig.~\ref{fig:avfids} in more detail reveals three distinct regions. For very small values of $\Delta$ (i.e., in the strong coupling regime) the limits predicted in \cite{dariano02} are recovered, yielding an average fidelity of $2/3$ for $N=1$ and $2$ and $\mathcal{F}_{av} = 7/10$ for $N=3$ and $N = 4$, respectively. On the other extreme, for $\Delta > 1$, we see that the average fidelity starts to decline rapidly. This can be understood if one notices that after a certain value of $\Delta$ the coupling is so weak that the procedure is essentially equivalent to randomly guessing, therefore yielding an average fidelity of $1/2$. The intermediate region is particularly intriguing because in the case of two and more spins the average fidelity is superior to what can be achieved when $\Delta \rightarrow 0$. This means that in this case a lesser coupling strength can achieve better results than a strong coupling.

\section{Large number of spins}
\label{sec:large-number-spins}
  
For $N>4$ the full calculation becomes cumbersome without approximations. To continue with larger $N$, we calculate a lower bound on $\mathcal{F}_{\mathrm{av}}$. For this, we resolve the identity $\mathbbm{1} = \left| \vec{r} \right\rangle\!\left\langle \vec{r}\right| ^{\otimes N} + (\mathbbm{1}-\left| \vec{r} \right\rangle\!\left\langle \vec{r}\right| ^{\otimes N})$ and insert it in the expression for $p(\vec{r})$. The reason for this choice is that $E(\vec{r})$ is diagonal in the eigenbasis of $\vec{r}\cdot \vec{S}$ and the Kraus operator leading to the optimal $\mathcal{F}_{\mathrm{av}}$ should have a large overlap with $\left| \vec{r} \right\rangle\!\left\langle \vec{r}\right| ^{\otimes N}$ \footnote{M.-O.~Renou \textit{et al.}, \textit{in prep}.}. In the following, we bound $p(\vec{r})  \geq | \left\langle \vec{r} \right| ^{\otimes N} E(\vec{r}) \left|\uparrow \right\rangle|^{2N}$. 

For general $N$, one has to find further simplifications for $p(\vec{r})$. For an $\mathcal{F}_{\mathrm{av}}$ scaling as $1-O(1/N)$ (i.e., the optimal scaling), it is necessary that $\left| \left\langle \vec{r} \right| \left. \uparrow \right\rangle \right|^{2} = 1-O(1/N)$ for almost all $\vec{r}$ which have $O(1)$ support from $p(\vec{r})$. Hence, the most important $\left| \vec{r} \right\rangle $ are $O(1/\sqrt{N})$-close to $\left|\uparrow \right\rangle $. Inspecting the matrix element $E_{\vec{r}} = \left\langle \vec{r} \right| ^{\otimes N} E(\vec{r}) \left| \uparrow \right\rangle ^{\otimes N}$ in the momentum basis of the pointer,
\begin{equation}
\label{eq:12}
E_{\vec{r}} \propto \int d \vec{p} e^{i \vec{r}\cdot \vec{p}} e^{- \Delta^2 p^2} \left\langle \vec{r} \right| ^{\otimes N} e^{-i \vec{p}\cdot \vec{S}} \left| \uparrow \right\rangle ^{\otimes N},
\end{equation}
one notices that this implies $\vec{p} = O(1/\sqrt{N})$, which in turn means that $\Delta = O(\sqrt{N})$. In this regime, it is advantageous to perform a Holstein-Primakoff transformation \cite{holstein40} to express $S_{+} = S_x + i S_y$ as $S_{+}=\sqrt{1-a^{\dagger}a/N} a$ with $[a^{\dagger},a] = 1$; similarly for $S_{-}=S_{+}^{\dagger}$ and \mbox{$S_z = \frac{1}{2} [S_{+},S_{-}] = \frac{N}{2}- a^{\dagger} a$}. The input state $\left| \uparrow \right\rangle ^{\otimes N}$ is transformed to the vacuum $| 0 \rangle $, which is an eigenstate of $a$ with eigenvalue 0. The state $| r \rangle ^{\otimes N}$ corresponds to a coherent state, since rotations on the Bloch sphere are approximated by planer displacements. In summary, we have the mapping $\left\langle \vec{r} \right| ^{\otimes N} e^{-i \vec{p}\cdot \vec{S}} \left|\uparrow \right\rangle ^{\otimes N} \rightarrow \left\langle 0 \right| \exp[f(a,a^{\dagger})] \left| 0 \right\rangle $ with a lengthy function $f$ in $a$ and $a^{\dagger}$.

The next step is to find the optimal value of $\Delta$. For this, we write $\exp[f(a,a^{\dagger})]$ in a Taylor series for $N\rightarrow \infty$. Keeping terms up to order $O(1)$, one can analytically calculate $\mathcal{F}_{\mathrm{av}}$ and find a maximum for $\Delta_{\mathrm{opt}} = \sqrt{N/8}$. Heuristic arguments indicate that the following results are asymptotically unchanged by adding $o(\sqrt{N})$ terms to $\Delta_{\mathrm{opt}}$.

\begin{figure}[htbp]
  \centerline{\includegraphics[width=\columnwidth]{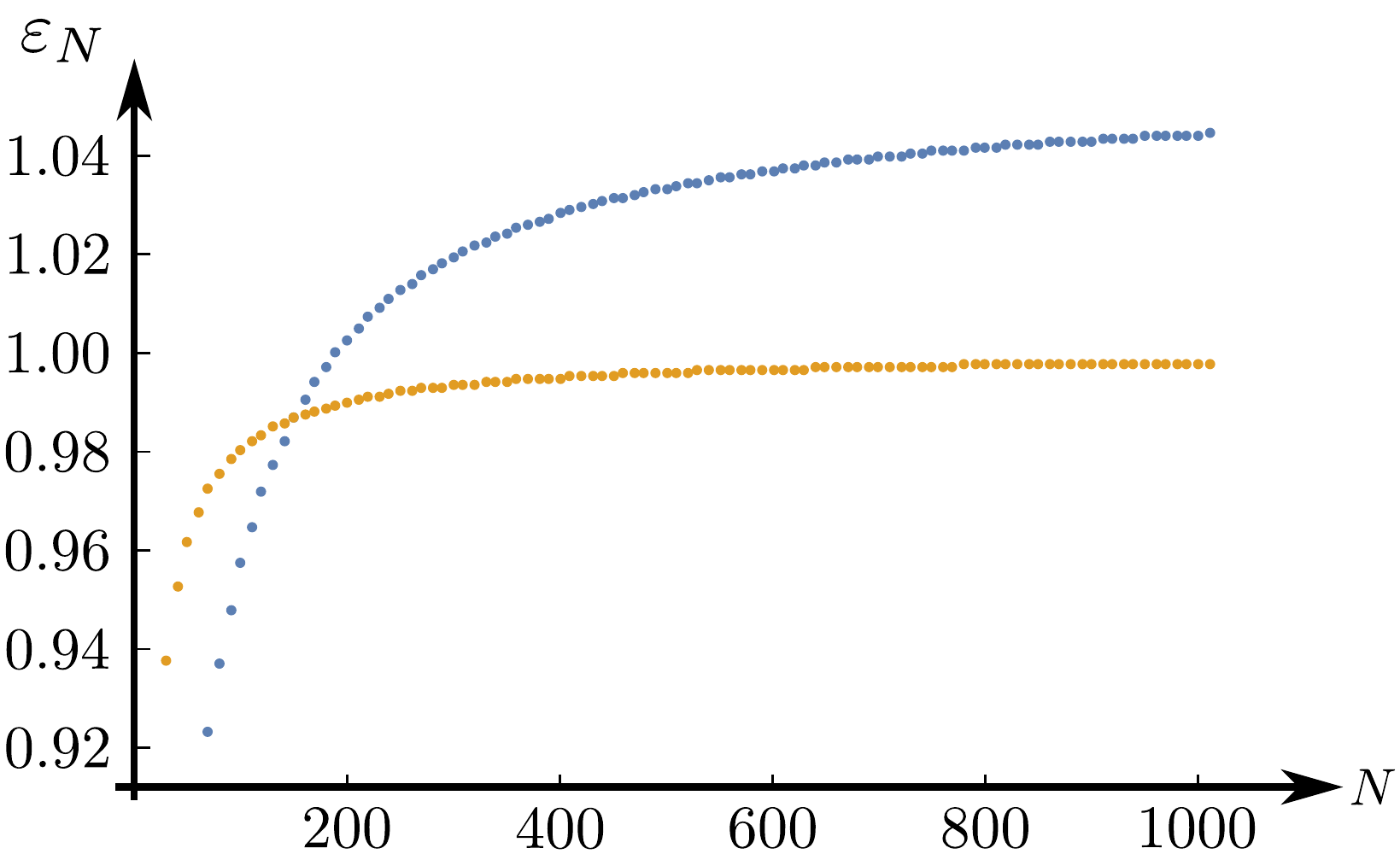}}
  \caption[]{\label{fig:scaling} Scaling factor $\epsilon_N \lesssim N(1-\mathcal{F}_{\mathrm{av}})$ from the approximate lower bound on $\mathcal{F}_{\mathrm{av}}$ (upper, blue curve) compared to the optimal scaling factor $N(1-\mathcal{F}_{\mathrm{opt}}) = (1+2/N)^{-1}$ (lower, orange curve). For large $N$, $\epsilon_N$ seems to stay slightly above one, which indicates a nonoptimal lower bound. For $N < 150$, the approximation clearly does not hold.}
\end{figure}

In order to have more precise results for $\mathcal{F}_{\mathrm{av}}$, we take into account terms up to order $O(1/N)$ in the power series of $\exp[f(a,a^{\dagger})]$. We are still able to analytically integrate over $\vec{p}$ in Eq.~(\ref{eq:12}) and the radial part of $\vec{r}$ in Eq.~\eqref{eq:11}. The final integration over $\theta$ is done numerically. We find that $\mathcal{F}_{\mathrm{av}} \gtrsim 1 - \epsilon_N / N$, where $\epsilon_N$ seems to asymptotically converge to a value slightly above one (see Fig.~\ref{fig:scaling}). This indicates that the approximate lower bound does asymptotically not coincide with $\mathcal{F}_{\mathrm{opt}} = 1-1/N + O(N^{-2})$. However, the relative difference between the two clearly goes to zero. Moreover, taking into account more matrix elements in addition to Eq.~(\ref{eq:12}) could reduce the gap.

\section{Measurement-induced disturbance}
\label{sec:meas-induc-dist}
Interestingly, it turns out that the presented measurement scheme is hardly invasive. To show this, we calculate the quantum fidelity between the pre- and post-measured state averaged over all measurement outcomes. The post-measured state reads $\rho_{\mathrm{post}} = \mathrm{Tr}_{\text{pointer}} U \left| \phi \right\rangle\!\left\langle \phi\right|  \otimes \left| \uparrow \right\rangle\!\left\langle \uparrow\right| ^{\otimes N} U^{\dagger}$. The disturbance is then defined to be $D = 1 - \left\langle \uparrow\right| ^{\otimes N} \rho_{\mathrm{post}} \left| \uparrow \right\rangle ^{\otimes N}$. For calculating the trace of $\rho_{\mathrm{post}}$ it is most convenient to work in the momentum basis of the pointer space. Then, simple manipulations give 
\begin{equation}
\label{eq:1}
D = 1- \int d^3p \left| \phi(\vec{p}) \right|^2 \left[ 1 - \sin^2(p/2) \sin^2 \theta \right]^N.
\end{equation}
For Gaussian pointers, this can in principle be calculated for any $N$.

\begin{figure}[htbp]
\centerline{\includegraphics[width=\columnwidth]{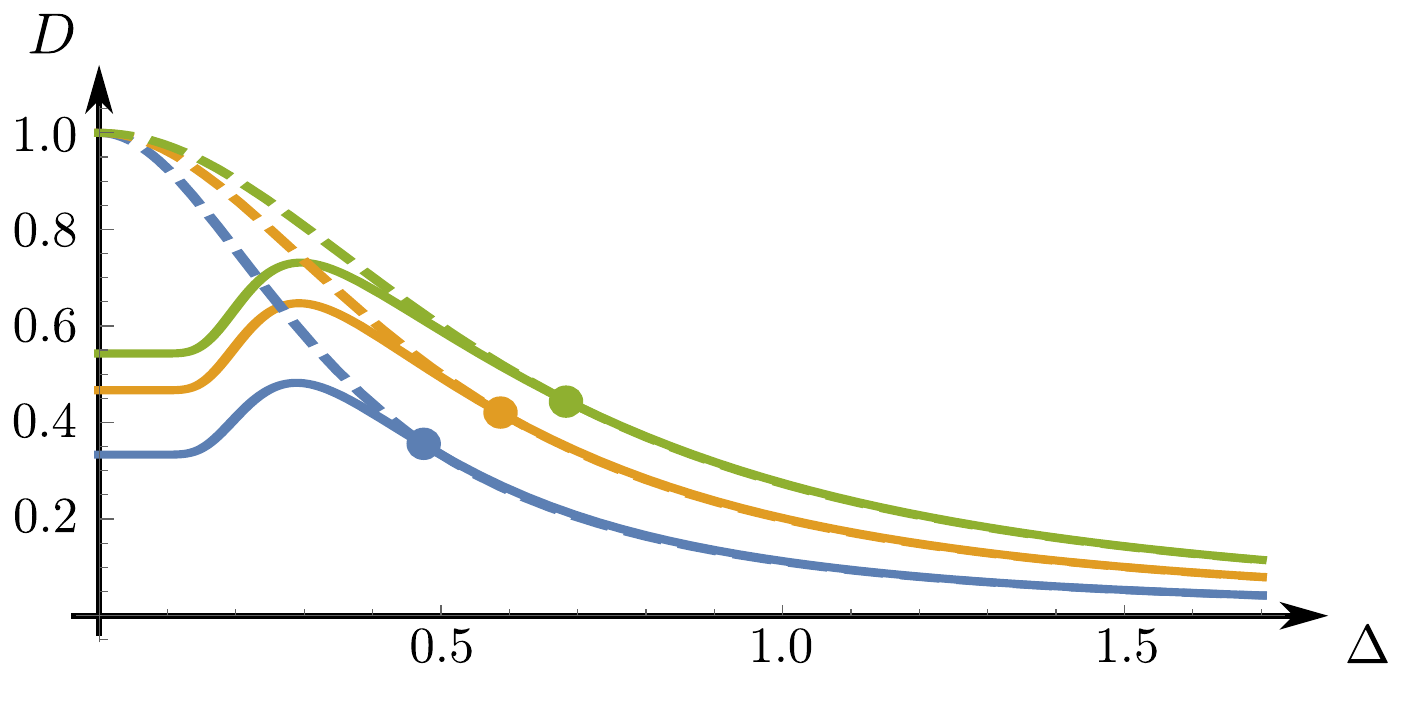}}
\caption[]{\label{fig:Disturbance} Disturbance measured by the quantum fidelity between the pre- and post-measured states as function of $\Delta$ for $N = 1,2,3$ (from bottom to top). The solid lines correspond to the exact expression, Eq.~(\ref{eq:1}), while the dashed curves are the lowest-order approximations for large $N$, $D \approx\left( 1+ 8\Delta^2/N \right)^{-1}$. The points on the curves indicate where $\Delta = \Delta_{\mathrm{opt}}$.}
\end{figure}

For small $N$, we observe that the maximal $D$ is found for some $\Delta>0$, which does however not correspond to the $\Delta$ leading to the maximal information (see Fig.~\ref{fig:Disturbance}).

For $N \gg 1$, a closed formula is only found for an approximated integrand. Similar as before, we set $\Delta = c \sqrt{N/8}$ and do a Taylor series for $N \rightarrow \infty$. In lowest order, one finds the Lorentzian-like function $D = (1+c^2)^{-1}$, which does not indicate any particular behavior at $c = 1$, for which $\mathcal{F}_{\mathrm{av}}$ is maximal in the large $N$ limit. Higher orders exhibit simple, but long expressions in $c$, which simplify for $c = 1$ to
\begin{equation}
\label{eq:2}
D = \frac{1}{2} + \frac{23}{1440 N^2} + O \left( \frac{1}{N^3} \right).
\end{equation}
Hence, in the limit of large $N$, the disturbance is close to the minimal disturbance $D_{\min} = (N+1)/(2N +1)$ in the case of optimal guessing fidelity \cite{Sacchi_Information-disturbance_2007}. A nonunit value seems to indicate a rather severe change in the state. We note, however, that the overlap between two states $| \vec{u} \rangle ^{\otimes N}$ and $| \vec{w} \rangle ^{\otimes N}$ generally decays exponentially in $N$ unless the angle between $\vec{u}$ and $\vec{w}$ is at most $O(1/\sqrt{N})$. A constant overlap for large $N$ indicates an $O(1/\sqrt{N})$ closeness between $\left| \uparrow \right\rangle^{\otimes N} $ and $\rho_{\mathrm{post}}$ on the Bloch sphere.

This becomes even more evident when we consider the Bloch vector of the spin ensemble before and after measurement. The expectation value $\langle \vec{S} \rangle$ for both states can be calculated in a straightforward manner. One has $\langle S_x \rangle = \langle S_y \rangle = 0$ for both $\left|\uparrow \right\rangle ^{\otimes N}$ and $\rho_{\mathrm{post}}$. For the $z$ component, we find $\langle S_z \rangle_{\left| \uparrow \right\rangle ^{\otimes N}} = N/2$ and $\langle S_z \rangle_{\rho_{\mathrm{post}}} = \frac{1}{6}N\left[ 1 + e^{-1/(8\Delta^2)}\left( 2- 1/(2 \Delta^2) \right) \right] = N/2 - 1 + O(1/N)$, where the last expression holds for $\Delta = \Delta_{\mathrm{opt}}$. In words, the length of the post-measured Bloch vector is only minimally reduced. A repetition of the measurement will give almost the same information about the Bloch vector than the first one. This is in stark contrast to some other optimal measurement strategies. For example, in Ref.~\cite{bagan05}, the single-spin measurements in a random direction leaves behind a completely depolarized product state; therefore making it impossible for a different pointer to gain any information about the initial state in a subsequent measurement.

\section{Discussion and open questions}
\label{sec:open-questions.-}

In summary, the simultaneous measurement of the noncommuting observables $S_x, S_y$ and $S_z$ can be made highly sensitive and hardly disturbing in a regime where the coupling to the pointers is relatively weak. This holds true in particular for large spin ensembles. Via an approximate Trotter expansion, our results also guide a quantitative analysis of proposals \cite{Poulin_Macroscopic_2005} for sequential measurements of noncommuting observables in the weak coupling regime. 

Given the features of the model together with a fairly simple scheme, the discussed model is reminiscent of a classical measurement. This observation allows us to speculate about a general class of ``macroscopic quantum measurements'' as measurements within the quantum formalism with high information gain, low disturbance and simple physical pointer models. In other words, the present model could be a starting point to understand the emergence of classical behaviour of quantum measurements for large systems. It might be interesting to study a connection to recent work on classifying different levels of simultaneous measurements \cite{Heinosaari_Simultaneous_2016}. Further research will be dedicated to other schemes of simultaneous measurements as a function of coupling strength, in particular when the observables have well defined macroscopic limits such as position and momentum.

It is likely that the very same measurement model can also be used to illustrate the classical appearance of macroscopic quantum states by changing the input state. Suppose a superposition of macroscopically distinct states (i.e., a Schr\"{o}dinger-cat state) \cite{leggett81} is perfectly prepared. We expect that the coherence between these distinct states cannot be witnessed if the system is subject to a coupling as in Eq.~\eqref{ham}. This reflects the basic idea of einselection \cite{zurek03}, where the measurement apparatus takes the role of the environment. In addition, it would be interesting to see whether the presented coupling could also serve as an illustrating example of quantum Darwinism \cite{zurek09}.

Further open questions include a better characterization of $\mathcal{F}_{\mathrm{av}}$ for $N\gg 1$ in terms of the precise scaling with $N$ as well as the robustness with respect to variations of $\Delta_{\mathrm{opt}}$. Also, more realistic initial states, that is, ones that better match the properties of real-world examples, are interesting to study. A first attempt would include nonmaximally polarized and thermal states. 

\textit{Acknowledgments.---} We thank Pavel Sekatski and Serge Massar for stimulating discussions. This work was supported by the National Swiss Science Foundation (SNSF), the Austrian Science Fund (FWF), grant number J3462, the COST Action No.~MP1006 and the European Research Council (ERC MEC).

\bibliographystyle{apsrev4-1}
\bibliography{../WeakMeasurement}

\begin{thebibliography}{18}%
\makeatletter
\providecommand \@ifxundefined [1]{%
 \@ifx{#1\undefined}
}%
\providecommand \@ifnum [1]{%
 \ifnum #1\expandafter \@firstoftwo
 \else \expandafter \@secondoftwo
 \fi
}%
\providecommand \@ifx [1]{%
 \ifx #1\expandafter \@firstoftwo
 \else \expandafter \@secondoftwo
 \fi
}%
\providecommand \natexlab [1]{#1}%
\providecommand \enquote  [1]{``#1''}%
\providecommand \bibnamefont  [1]{#1}%
\providecommand \bibfnamefont [1]{#1}%
\providecommand \citenamefont [1]{#1}%
\providecommand \href@noop [0]{\@secondoftwo}%
\providecommand \href [0]{\begingroup \@sanitize@url \@href}%
\providecommand \@href[1]{\@@startlink{#1}\@@href}%
\providecommand \@@href[1]{\endgroup#1\@@endlink}%
\providecommand \@sanitize@url [0]{\catcode `\\12\catcode `\$12\catcode
  `\&12\catcode `\#12\catcode `\^12\catcode `\_12\catcode `\%12\relax}%
\providecommand \@@startlink[1]{}%
\providecommand \@@endlink[0]{}%
\providecommand \url  [0]{\begingroup\@sanitize@url \@url }%
\providecommand \@url [1]{\endgroup\@href {#1}{\urlprefix }}%
\providecommand \urlprefix  [0]{URL }%
\providecommand \Eprint [0]{\href }%
\providecommand \doibase [0]{http://dx.doi.org/}%
\providecommand \selectlanguage [0]{\@gobble}%
\providecommand \bibinfo  [0]{\@secondoftwo}%
\providecommand \bibfield  [0]{\@secondoftwo}%
\providecommand \translation [1]{[#1]}%
\providecommand \BibitemOpen [0]{}%
\providecommand \bibitemStop [0]{}%
\providecommand \bibitemNoStop [0]{.\EOS\space}%
\providecommand \EOS [0]{\spacefactor3000\relax}%
\providecommand \BibitemShut  [1]{\csname bibitem#1\endcsname}%
\let\auto@bib@innerbib\@empty
%</preamble>
\bibitem [{\citenamefont {Massar}\ and\ \citenamefont
  {Popescu}(1995)}]{massar95}%
  \BibitemOpen
  \bibfield  {author} {\bibinfo {author} {\bibfnamefont {S.}~\bibnamefont
  {Massar}}\ and\ \bibinfo {author} {\bibfnamefont {S.}~\bibnamefont
  {Popescu}},\ }\href {\doibase 10.1103/PhysRevLett.74.1259} {\bibfield
  {journal} {\bibinfo  {journal} {Phys. Rev. Lett.}\ }\textbf {\bibinfo
  {volume} {74}},\ \bibinfo {pages} {1259} (\bibinfo {year}
  {1995})}\BibitemShut {NoStop}%
\bibitem [{\citenamefont {Bagan}\ \emph {et~al.}(2005)\citenamefont {Bagan},
  \citenamefont {Monras},\ and\ \citenamefont {Mu\~noz{-}Tapia}}]{bagan05}%
  \BibitemOpen
  \bibfield  {author} {\bibinfo {author} {\bibfnamefont {E.}~\bibnamefont
  {Bagan}}, \bibinfo {author} {\bibfnamefont {A.}~\bibnamefont {Monras}}, \
  and\ \bibinfo {author} {\bibfnamefont {R.}~\bibnamefont {Mu\~noz{-}Tapia}},\
  }\href {\doibase 10.1103/PhysRevA.71.062318} {\bibfield  {journal} {\bibinfo
  {journal} {Phys. Rev. A}\ }\textbf {\bibinfo {volume} {71}},\ \bibinfo
  {pages} {062318} (\bibinfo {year} {2005})}\BibitemShut {NoStop}%
\bibitem [{\citenamefont {D'Ariano}\ \emph {et~al.}(2002)\citenamefont
  {D'Ariano}, \citenamefont {Presti},\ and\ \citenamefont
  {Sacchi}}]{dariano02}%
  \BibitemOpen
  \bibfield  {author} {\bibinfo {author} {\bibfnamefont {G.}~\bibnamefont
  {D'Ariano}}, \bibinfo {author} {\bibfnamefont {P.~L.}\ \bibnamefont
  {Presti}}, \ and\ \bibinfo {author} {\bibfnamefont {M.}~\bibnamefont
  {Sacchi}},\ }\href {\doibase http://dx.doi.org/10.1016/S0375-9601(01)00809-X}
  {\bibfield  {journal} {\bibinfo  {journal} {Physics Letters A}\ }\textbf
  {\bibinfo {volume} {292}},\ \bibinfo {pages} {233 } (\bibinfo {year}
  {2002})}\BibitemShut {NoStop}%
\bibitem [{\citenamefont {Arthurs}\ and\ \citenamefont
  {Kelly}(1965)}]{arthurs65}%
  \BibitemOpen
  \bibfield  {author} {\bibinfo {author} {\bibfnamefont {E.}~\bibnamefont
  {Arthurs}}\ and\ \bibinfo {author} {\bibfnamefont {J.~L.}\ \bibnamefont
  {Kelly}},\ }\href {\doibase 10.1002/j.1538-7305.1965.tb01684.x} {\bibfield
  {journal} {\bibinfo  {journal} {Bell System Technical Journal}\ }\textbf
  {\bibinfo {volume} {44}},\ \bibinfo {pages} {725} (\bibinfo {year}
  {1965})}\BibitemShut {NoStop}%
\bibitem [{Note1()}]{Note1}%
  \BibitemOpen
  \bibinfo {note} {The average fidelity of 3/4 is exactly the same value as
  estimating a classical spin with a Stern-Gerlach-type experiment with
  randomly chosen measurement axis.}\BibitemShut {Stop}%
\bibitem [{\citenamefont {Poulin}(2005)}]{Poulin_Macroscopic_2005}%
  \BibitemOpen
  \bibfield  {author} {\bibinfo {author} {\bibfnamefont {D.}~\bibnamefont
  {Poulin}},\ }\href@noop {} {\bibfield  {journal} {\bibinfo  {journal} {Phys.
  Rev. A}\ }\textbf {\bibinfo {volume} {71}},\ \bibinfo {pages} {022102}
  (\bibinfo {year} {2005})}\BibitemShut {NoStop}%
\bibitem [{\citenamefont {Bu\ss{}hardt}\ and\ \citenamefont
  {Freyberger}(2010)}]{Busshardt_Timing_2010}%
  \BibitemOpen
  \bibfield  {author} {\bibinfo {author} {\bibfnamefont {M.}~\bibnamefont
  {Bu\ss{}hardt}}\ and\ \bibinfo {author} {\bibfnamefont {M.}~\bibnamefont
  {Freyberger}},\ }\href@noop {} {\bibfield  {journal} {\bibinfo  {journal}
  {Phys. Rev. A}\ }\textbf {\bibinfo {volume} {82}},\ \bibinfo {pages} {042117}
  (\bibinfo {year} {2010})}\BibitemShut {NoStop}%
\bibitem [{\citenamefont {von Neumann}(1955)}]{neumann55}%
  \BibitemOpen
  \bibfield  {author} {\bibinfo {author} {\bibfnamefont {J.}~\bibnamefont {von
  Neumann}},\ }\href@noop {} {\emph {\bibinfo {title} {Mathematical Foundations
  of Quantum Mechanics}}}\ (\bibinfo  {publisher} {Princeton University
  Press},\ \bibinfo {address} {Princeton, New Jersey},\ \bibinfo {year}
  {1955})\BibitemShut {NoStop}%
\bibitem [{\citenamefont {Busch}\ \emph {et~al.}(1991)\citenamefont {Busch},
  \citenamefont {Lahti},\ and\ \citenamefont
  {Mittelstaedt}}]{Busch_Quantum_1991}%
  \BibitemOpen
  \bibfield  {author} {\bibinfo {author} {\bibfnamefont {P.}~\bibnamefont
  {Busch}}, \bibinfo {author} {\bibfnamefont {P.~J.}\ \bibnamefont {Lahti}}, \
  and\ \bibinfo {author} {\bibfnamefont {P.}~\bibnamefont {Mittelstaedt}},\
  }in\ \href {\doibase 10.1007/978-3-662-13844-1_3} {\emph {\bibinfo
  {booktitle} {The {{Quantum Theory}} of {{Measurement}}}}},\ \bibinfo {series
  and number} {\bibinfo {series} {Lecture Notes in Physics Monographs}\
  No.~\bibinfo {number} {2}}\ (\bibinfo  {publisher} {{Springer Berlin
  Heidelberg}},\ \bibinfo {year} {1991})\ pp.\ \bibinfo {pages}
  {27--98}\BibitemShut {NoStop}%
\bibitem [{\citenamefont {Peres}(2002)}]{peres02}%
  \BibitemOpen
  \bibfield  {author} {\bibinfo {author} {\bibfnamefont {A.}~\bibnamefont
  {Peres}},\ }\href@noop {} {\emph {\bibinfo {title} {Quantum Theory: Concepts
  and Methods}}}\ (\bibinfo  {publisher} {Kluwer Academic Publishers},\
  \bibinfo {address} {New York},\ \bibinfo {year} {2002})\BibitemShut {NoStop}%
\bibitem [{Note2()}]{Note2}%
  \BibitemOpen
  \bibinfo {note} {Note that this choice is not always optimal. For some
  (nonoptimal) values of $\Delta $, the optimal choice is $| \protect
  \mathaccentV {vec}17Ew_{\protect \mathaccentV {vec}17Er} \protect \rangle = |
  -\protect \mathaccentV {vec}17Er \protect \rangle $. For symmetry reasons the
  optimal guess has to lie on the axis of $\protect \mathaccentV
  {vec}17Er$.}\BibitemShut {Stop}%
\bibitem [{Note3()}]{Note3}%
  \BibitemOpen
  \bibinfo {note} {M.-O.~Renou \protect \textit {et al.}, \protect \textit {in
  prep}.}\BibitemShut {Stop}%
\bibitem [{\citenamefont {Holstein}\ and\ \citenamefont
  {Primakoff}(1940)}]{holstein40}%
  \BibitemOpen
  \bibfield  {author} {\bibinfo {author} {\bibfnamefont {T.}~\bibnamefont
  {Holstein}}\ and\ \bibinfo {author} {\bibfnamefont {H.}~\bibnamefont
  {Primakoff}},\ }\href {\doibase 10.1103/PhysRev.58.1098} {\bibfield
  {journal} {\bibinfo  {journal} {Phys. Rev.}\ }\textbf {\bibinfo {volume}
  {58}},\ \bibinfo {pages} {1098} (\bibinfo {year} {1940})}\BibitemShut
  {NoStop}%
\bibitem [{\citenamefont {Sacchi}(2007)}]{Sacchi_Information-disturbance_2007}%
  \BibitemOpen
  \bibfield  {author} {\bibinfo {author} {\bibfnamefont {M.~F.}\ \bibnamefont
  {Sacchi}},\ }\href@noop {} {\bibfield  {journal} {\bibinfo  {journal} {Phys.
  Rev. A}\ }\textbf {\bibinfo {volume} {75}},\ \bibinfo {pages} {012306}
  (\bibinfo {year} {2007})}\BibitemShut {NoStop}%
\bibitem [{\citenamefont {Heinosaari}(2016)}]{Heinosaari_Simultaneous_2016}%
  \BibitemOpen
  \bibfield  {author} {\bibinfo {author} {\bibfnamefont {T.}~\bibnamefont
  {Heinosaari}},\ }\href@noop {} {\bibfield  {journal} {\bibinfo  {journal}
  {Phys. Rev. A}\ }\textbf {\bibinfo {volume} {93}},\ \bibinfo {pages} {042118}
  (\bibinfo {year} {2016})}\BibitemShut {NoStop}%
\bibitem [{\citenamefont {Caldeira}\ and\ \citenamefont
  {Leggett}(1981)}]{leggett81}%
  \BibitemOpen
  \bibfield  {author} {\bibinfo {author} {\bibfnamefont {A.~O.}\ \bibnamefont
  {Caldeira}}\ and\ \bibinfo {author} {\bibfnamefont {A.~J.}\ \bibnamefont
  {Leggett}},\ }\href {\doibase 10.1103/PhysRevLett.46.211} {\bibfield
  {journal} {\bibinfo  {journal} {Phys. Rev. Lett.}\ }\textbf {\bibinfo
  {volume} {46}},\ \bibinfo {pages} {211} (\bibinfo {year} {1981})}\BibitemShut
  {NoStop}%
\bibitem [{\citenamefont {Zurek}(2003)}]{zurek03}%
  \BibitemOpen
  \bibfield  {author} {\bibinfo {author} {\bibfnamefont {W.~H.}\ \bibnamefont
  {Zurek}},\ }\href {\doibase 10.1103/RevModPhys.75.715} {\bibfield  {journal}
  {\bibinfo  {journal} {Rev. Mod. Phys.}\ }\textbf {\bibinfo {volume} {75}},\
  \bibinfo {pages} {715} (\bibinfo {year} {2003})}\BibitemShut {NoStop}%
\bibitem [{\citenamefont {Zurek}(2009)}]{zurek09}%
  \BibitemOpen
  \bibfield  {author} {\bibinfo {author} {\bibfnamefont {W.~H.}\ \bibnamefont
  {Zurek}},\ }\href {\doibase 10.1038/nphys1202} {\bibfield  {journal}
  {\bibinfo  {journal} {Nat Phys}\ }\textbf {\bibinfo {volume} {5}},\ \bibinfo
  {pages} {181} (\bibinfo {year} {2009})}\BibitemShut {NoStop}%
\end{thebibliography}%
 
\end{document}